# Power-Performance Trade-offs in Nanometer-Scale Multi-Level Caches Considering Total Leakage

Robert Bai[1], Nam-Sung Kim[2], Tae Ho Kgil[1], Dennis Sylvester[1], Trevor Mudge[1]
[1] University of Michigan, EECS Department, Ann Arbor, MI 48109 {rbai, tkgil, dennis, tnm} @eecs.umich.edu
[2] Intel Corporation, Portland, Oregon, nam.sung.kim@intel.com

## Abstract

In this paper, we investigate the impact of $T_{ox}$ and $V_{th}$ on power performance trade-offs for on-chip caches. We start by examining the optimization of the various components of a single level cache and then extend this to two level cache systems. In addition to leakage, our studies also account for the dynamic power expended as a result of cache misses. Our results show that one can often reduce overall power by increasing the size of the L2 cache if we only allow one pair of $V_{th}/T_{ox}$ in L2. However, if we allow the memory cells and the peripherals to have their own $V_{th}$'s and $T_{ox}$'s, we show that a two-level cache system with smaller L2's will yield less total leakage. We further show that two $V_{th}$'s and two $T_{ox}$'s are sufficient to get close to an optimal solution, and that $V_{th}$ is generally a better design knob than $T_{ox}$ for leakage optimization, thus it is better to restrict the number of $T_{ox}$'s rather than $V_{th}$'s if cost is a concern.

## 1. Introduction

Leakage power is a problem for all microprocessor circuit components, but it is a particularly important problem in processor on-chip caches where a large number of potentially high-leakage cross-coupled inverters — the storage elements of caches — are integrated in great numbers. We can expect the fraction of the leakage power to exceed that of the dynamic power in future processor generations. There have been several previous studies on cache leakage power reduction [1-7]; all of them focused on subthreshold leakage power. However, with aggressive $T_{ox}$ scaling, gate leakage power can potentially surpass the subthreshold leakage at low $T_{ox}$. In this paper, we investigate various techniques to minimize total (gate + subthreshold) leakage power plus dynamic power under delay constraints by systematically assigning values for $T_{ox}$ and $V_{th}$ for single cache, two-level caching system and an entire microprocessor memory system consisting of L1, L2 cache and main memory.

## 2. Circuit Evaluation Methodology

For our experiment, we have used the technology files from Berkeley Predictive Technology Model (BPTM) for a 65nm technology node [8]. We then characterize the technology files for a range of $V_{th}$ and $T_{ox}$ values. We let $V_{th}$ vary from 0.2V to 0.5V, while allowing $T_{ox}$ to scale from 10Å to 14Å. The lower limits of these ranges are chosen to reflect typical values of high-performance logic for the studied technology node. Such transistors would be required for the non-memory portion of a processor or system. While there is no physical reason for a $V_{th}$ upper bound, we expect that values above 0.5V are unlikely in 65nm technology with approximately 1V supply. The increase of $T_{ox}$ while maintaining the same drawn channel length may cause the gate terminal to lose control of the conduction state of the channel due to DIBL effect [9]. Hence, when $T_{ox}$ changes, the drawn channel length must be scaled appropriately. Also in order to maintain memory cell stability, the widths of the transistors in the memory cell need to be adjusted proportionately with the change in the drawn channel lengths. Thus the impact of $T_{ox}$ scaling on the cell area must be taken into account, as the cell will grow in both horizontal and vertical dimensions.

## 3. Delay and Leakage Power Models

First we have re-designed the cache netlists used in [7] to target for 65nm technology node. We assume that internally, the cache consists of four components: memory cell array and sense amplifier, decoder, address bus drivers, and data bus drivers. Second, it is observed through extensive HSPICE simulation that the total leakage current of memory cell array is exponentially dependent on $T_{ox}$ and $V_{th}$. We then approximate the total leakage power as follows:

$$P_{total}(V_{th}, T_{ox}) = A_0 + A_1 * e^{a_1 * V_{th}} + A_2 * e^{a_2 * T_{ox}}$$

On the other hand, the delay of the array is shown to be linear with $T_{ox}$ and over the range of our interest its dependence on $V_{th}$ can be approximated to an exponential growth function with very small exponents as follows:

$$T_d(V_{th}, T_{ox}) = k_0 + k_1 * e^{(k_3 * V_{th})} + k_2 * T_{ox}$$

Although these total leakage and delay trends are for the memory cell array, the same trends also hold for the rest of cache memory components — decoders and address/data bus drivers. Therefore, we can model the total leakage and delay of each component in the same way as we do for the memory cell array assuming that both total leakage and delay of each component are independent from one another. Thus we can approximate both the total leakage and the delay of a cache system by summing up the leakage and delay of each cache component.

## 4. Single Cache Leakage Optimization

To examine the dependence of leakage power on $V_{th}$ and $T_{ox}$ assignment, we study three different $V_{th}/T_{ox}$ assignment schemes:

- Scheme I: assign independent $V_{th}$'s and $T_{ox}$'s to each cache component.
- Scheme II: assign a $V_{th}/T_{ox}$ pair to the memory cell array and another pair to the remaining three cache components.
- Scheme III: assign the same $V_{th}/T_{ox}$ pair to all four cache components.

We formulate the problem of minimizing the leakage power given the delay constraint as the following optimization problem [10]:

***Minimize*** $LeakagePower(V_{th1}, T_{ox1}, ... V_{th4}, T_{ox4})$
$= A_0 + A_1 * e^{a_1 * V_{th1}} + A_2 * e^{a_2 * T_{ox1}} + ... + A_7 * e^{a_7 * V_{th4}} + A_8 * e^{a_8 * T_{ox4}}$
***Subject to*** $T_d(V_{th1}, T_{ox1}, ..., V_{th4}, T_{ox4}) =$
$B_0 + B_1 * e^{(b_1 * V_{th1})} + B_2 * T_{ox1} + ... + B_7 * e^{(b_4 * V_{th4})} + B_8 * T_{ox4};$
$10\text{ Å} \le T_{ox1}, T_{ox2}, T_{ox3}, T_{ox4} \le 14\text{ Å};$
$0.2V \le V_{th1}, V_{th2}, V_{th3}, V_{th4} \le 0.5V;$

In our optimization process, we have chosen $V_{th}$ and $T_{ox}$ to take on discrete values with small step size. The optimization shows scheme III is the worst performer, and scheme I is the best. However, scheme II is only slightly behind scheme I for the same delay constraint, but from a process standpoint, scheme I is more costly than scheme II. Therefore, it is the preferred scheme, as it is not only economically feasible but also achieves close to optimal leakage. It is worth noting that in schemes I and II, high values of $V_{th}$ and thick $T_{ox}$'s are always assigned to the memory cell arrays, and $V_{th}/T_{ox}$ in the peripheral components have been set sufficiently low to help meet the delay target. To gain further insight into the selection of the decision variables during the optimization process, we perform an experiment in which for a 16KB cache we hold either $V_{th}$ or $T_{ox}$ constant, and at the same time observe how leakage power is impacted by the other decision variable independently. In Figure 1, we show four curves, two of which are constructed by fixing $T_{ox}$ at 10Å and 14Å, respectively, and the other two are created by fixing $V_{th}$ at 0.2V and 0.4V, respectively. It is evident that the leakage is more sensitive to $T_{ox}$ than $V_{th}$, and the delay doesn't show as wide a range when $V_{th}$ is fixed as when $T_{ox}$ is fixed. Hence, to achieve minimum overall leakage, it is best to set $T_{ox}$ conservatively at a high value and let $V_{th}$ be the knob designers can vary to meet a delay constraint.

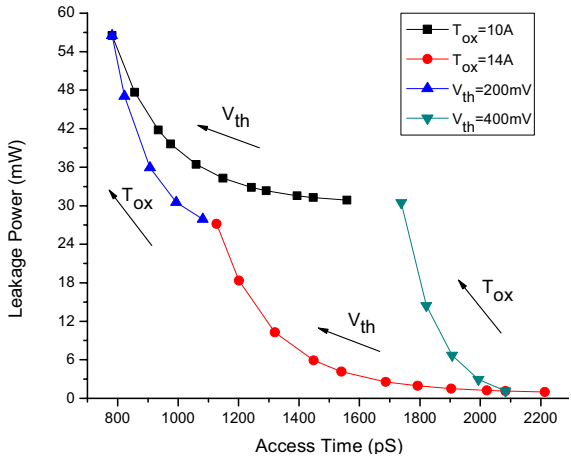

**Figure 1. Fixed $V_{th}$ vs. Fixed $T_{ox}$**

## 5. Two-Level Cache Leakage Optimization

We use architectural simulations to gather cache access statistics for each L1 and L2 cache size combination. To perform our evaluation, results from various benchmark suites such as SPEC2000, SPECWEB, TPC/C, etc., are collected.

*L2 Cache Leakage Power Optimization.* Due to its size, L2 caches naturally consume much more leakage than L1 caches. In the first of our experiments, we fix the size of an L1 cache and assign the default $V_{th}$ and $T_{ox}$ to the L1 cache, and then proceed to see which L2 organization would yield better leakage whilst still meeting the same average memory access time (AMAT) constraint of the two-level cache system. For example, we can adjust both $V_{th}$ and $T_{ox}$ knobs to make two different size caches that have the same AMAT: note that the AMAT is a function of both the cache miss rate and access (hit) time. The results show that generally the bigger L2 consumes less leakage power than smaller ones under the same delay constraint. This agrees with the trends presented in [7] that focused only on the $V_{th}$ assignment to optimize L2 cache leakage power. A larger L2 cache results in a smaller miss rate and faster AMAT, therefore $V_{th}$ and $T_{ox}$ of L2 can be set more conservatively for a larger L2 than for a smaller one. Nevertheless, having the largest available L2 does not always yield the best leakage. This is because we reach a point where the leakage of a very large L2 outweighs the benefit of the improvement in the L2 miss rate.

In the second part of our analysis, we assign a $V_{th}/T_{ox}$ pair to the core array cells in an L2 cache and another pair to its peripheral circuitry. In this scenario, there are two ways to improve AMAT. One is through reducing the L2 miss rate by employing a large L2 cache as was done in the first part of our analysis; the other is by setting the $V_{th}/T_{ox}$ assignments more aggressively in the peripheral circuitry. We found that the latter approach works better in the cases we have investigated. After optimization we see that $V_{th}$ and $T_{ox}$ in the core cell arrays are always set much more conservatively than those in the peripheral circuitry. This allows the leakiest component in the L2, which is the core cell array, to take on high values for $V_{th}$ and $T_{ox}$, thus saving leakage. At the same time, we can still meet the target delay because $V_{th}$ and $T_{ox}$ in the peripheral circuitry can be set sufficiently low.

*L1 Cache Leakage Power Optimization.* Local L1 cache miss rates are already very low and they do not vary much amongst the L1 caches ranging from 4K to 64K as illustrated in [7]. Hence given a fixed L2, the key to minimizing total leakage power is to reduce the leakage power consumed by L1. A smaller L1 will consume less leakage and at the same time a smaller L1 will be faster. Therefore, a small L1 will probably be the optimal solution

*Entire Processor Memory System Energy Optimization.* We determine the optimal number of $V_{th}$ and $T_{ox}$ values needed to achieve the optimal total energy for a system comprising of L1, L2 and main memory. Figure 2. shows that the best scheme has 2 $T_{ox}$'s and 3 $V_{th}$'s. However, the difference between a system with dual $T_{ox}$, dual $V_{th}$ and that with dual $T_{ox}$, triple $V_{th}$ is very small. So in general a process with dual $T_{ox}$ and dual $V_{th}$ is sufficient to achieve near optimal total energy. It is also worth noting that a single $T_{ox}$ and dual $V_{th}$ process outperforms that with a single $V_{th}$ and dual $T_{ox}$ for the same reason that we discovered in Section 4: $V_{th}$ is generally a more effective design knob than $T_{ox}$.

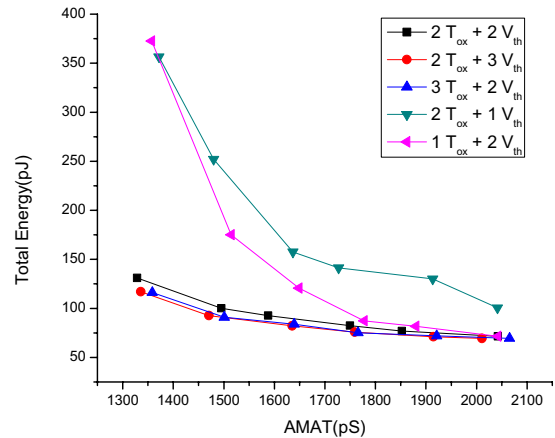

**Figure 2. ($T_{ox}$, $V_{th}$) Tuple Problem**